\documentclass[onecolumn,superscriptaddress,notitlepage,aps,pra,showkeys,10pt]{revtex4-1}                    
\usepackage{graphicx}
\usepackage{SIunits}
\begin{document}

\title{COMPARISON OF COULOMB BLOCKADE THERMOMETERS WITH THE INTERNATIONAL TEMPERATURE SCALE PLTS-2000}

\author{Matthias Meschke}
\affiliation{Low Temperature Laboratory, Aalto University, P.O. Box 15100, FI-00076 AALTO, Finland}
\author{Jost Engert}
\affiliation{Physikalisch-Technische Bundesanstalt, Abbestrasse 2-12, 10587 Berlin, Germany}
\author{Dieter Heyer}
\affiliation{Physikalisch-Technische Bundesanstalt, Abbestrasse 2-12, 10587 Berlin, Germany}
\author{Jukka P. Pekola}
\affiliation{Low Temperature Laboratory, Aalto University, P.O. Box 15100, FI-00076 AALTO, Finland}


\begin{abstract}
The operation of the primary Coulomb blockade thermometer (CBT) is based on a measurement of bias voltage dependent conductance of arrays of tunnel junctions between normal metal electrodes. Here we report on a comparison of a CBT with a high accuracy realization of the PLTS-2000 temperature scale in the range from 0.008 K to 0.65 K. An overall agreement of about 1\% was found for temperatures above 0.25 K. For lower temperatures increasing differences are caused by thermalization problems which are accounted for by numerical calculations based on electron-phonon decoupling.
\end{abstract}

\maketitle

\section{Introduction}

Traceable temperature measurements need an internationally accepted reference. For practical measurements of temperature in accord with the International System of Units (SI) international temperature scales are available - the International Temperature Scale of 1990 (ITS-90) \cite{Preston} and the Provisional Low Temperature Scale of 2000 (PLTS-2000) \cite{BIPM,Rusby}. The scale definitions and the supplementary documents issued by the Bureau International des Poids et Mesures (BIPM) \cite{BIPM2} provide detailed information on the realization of the ITS-90 and PLTS-2000 as well as the uncertainties which can be reached. However, recent developments in thermometry calling for a more flexible approach have resulted in the adoption of the ''Mise en pratique for the definition of the Kelvin'' by the Comit\'{e} Consultatif de Thermom\'{e}trie (CCT) of the BIPM. This document \cite{BIPM2} summarizes all information for practical temperature measurements in accord with the SI. Importantly, also primary thermometers for direct measurements of temperature now can be included in the ''Mise en pratique'' provided an assessment of the uncertainties for those measurements exists. 
Investigation of mesoscopic and nanoscopic objects has made large progress during the last decade. Along with an increased understanding of such structures, improved production technologies made
new classes of thermometers available \cite{Giazotto}. As an example, thermometers based on a tunnel junction or arrays of them as the Coulomb blockade thermometer (CBT) \cite{cbt94}, the single junction thermometer (SJT) \cite{PRL-Pekola08} or the shot noise thermometer (SNT) \cite{Spietz1,Spietz2} have been shown to work in a broad temperature range. The mentioned thermometers can be operated in primary mode because the measured quantity is in these devices directly linked to temperature without any need of a calibration against a temperature scale or reference. In combination with their simplicity of application these thermometers have the potential to replace a variety of secondary thermometers currently in use in the low temperature range. Nevertheless, a direct evaluation in terms of the international temperature scales is still lacking. Therefore, the main objective of this work is to investigate the CBT as a primary thermometer in the temperature range below 1 K by comparison with a high accuracy realization of the PLTS-2000.

CBTs are usually operated in the regime of weak Coulomb blockade \cite{cbt94} where the charging energy $E_{\rm{C}}$ is small compared to temperature, $E_{\rm{C}}$ $\ll$ $k_{\rm{B}}T$ ($k_{\rm{B}}$ - Boltzmann constant). Then the absolute temperature is obtained from the half width $V_{1/2}$ of the dip in the bias voltage dependent conductance curves according to

\begin{equation}
T_{\rm{CBT}}  = \frac{e V_{1/2}} {5.439Nk_{\rm B}},
\label{CBTnumcorr}
\end{equation}

where $\Delta G/G_{\rm T}$ is the depth of the normalized conductance dip, $N$ is the number of tunnel junctions in series, and $e$ is the electron charge. The equation holds also when the condition of small charging energy starts to fail when a linear correction term of $(1+0.4\Delta G/G_{\rm T})$ is introduced in the denominator of Eq. (\ref {CBTnumcorr}) \cite{JLTP-Farhanfar}. Alternatively the conductance curves can be calculated numerically without any approximation \cite{cbt94}, Eqs. (1)-(5). One advantage of the numerical calculation is that overheating effects due to small electron-phonon coupling can be included in the modeling as described in \cite{JLTP-Meschke}.
CBT has the advantage that many (in our case 100) junctions in series produce an enhanced signal, but the distribution in the junction resistances introduces errors in the absolute temperature readout. In the sensors reported in this work, we expect the fabrication inhomogeneity of the junction resistances to be below 10\% resulting in an uncertainty in temperature lower than 2\% \cite{JLTP-Farhanfar}. In addition to the conventional CBT thermometer, we present in this work a first comparison of a single junction thermometer (SJT) \cite{PRL-Pekola08}, that is not affected by fabrication inhomogeneity, with the PLTS-2000 scale. SJT demands on the other hand an enhanced experimental effort to precisely measure the smaller signal experimentally at low temperatures. 

\section{Fabrication parameters of CBT and SJT thermometers}
Both thermometers use aluminum oxide for the tunnel barriers because this material results in the highest quality junctions. However, aluminum is a superconducting material below $T$ $\approx$ 1.2 K and the working principles of both thermometers require normal metal islands. Consequently, a magnetic field on the order of 30 mT is required to suppress superconductivity in the sensors.

\begin{figure}[htbp]
  \centering
  \begin{minipage}[b]{0.39\textwidth}
    \includegraphics[width=\textwidth]{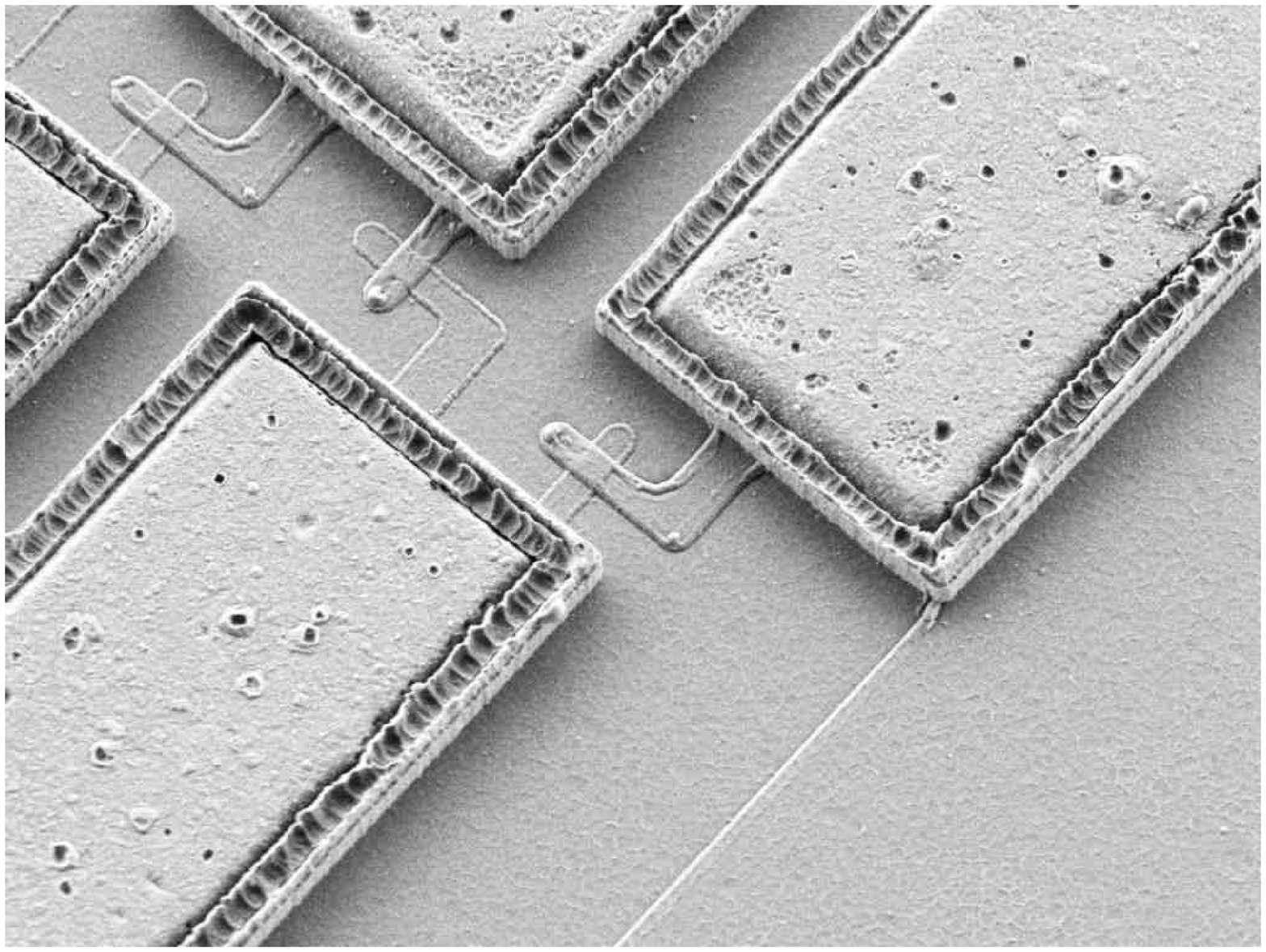}  
  \end{minipage}
  \begin{minipage}[b]{0.8\textwidth}
    \includegraphics[width=\textwidth]{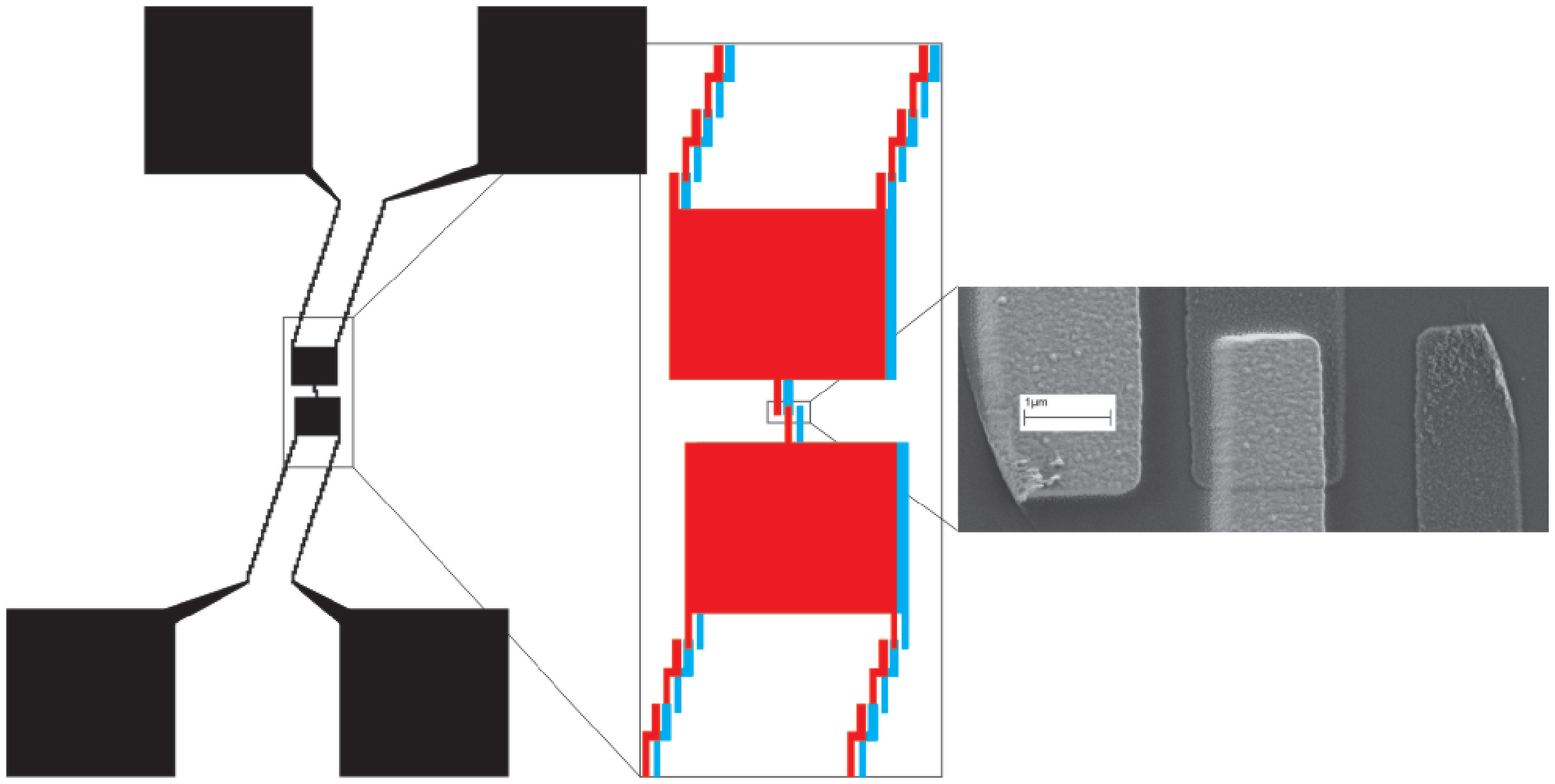}  
  \end{minipage}
  \caption{(top) Electron microscope image of three CBT junctions with a size of ($0.8\times0.8$) $\rm{\mu m^2}$ from one CBT sensor with attached thick cooling fins. Each island has additionally a big ($\approx 50$ $\rm{\mu m} \times20$ $\rm{\mu m}$), $2$ $\rm{\mu m}$ thick aluminum cooling fin attached to improve the electron thermalization.
(bottom)  Fabrication of the SJ thermometer: (left) the overall structure for the mask, written with electron beam lithography into a PMMA-MMA resist scheme \cite{Dolan}: four bonding pads connect via arrays of 20 tunnel junctions each to the central junction with additional large metal area for good thermalization. The metallization is done under two tilting angles (scheme in the middle) to produce the tunnel junctions: first 50 nm of aluminum is deposited, oxidized for 10 min in 100 mbar of oxygen and subsequently a second aluminum layer (red) forms the tunnel junction in the center (right SEM image) and in the connecting arrays. An additional 100 nm thick copper film on top of the 2nd layer of aluminum enhances electron thermalization. The area of the central tunnel junction is (1.8 x 1.2) $\mu\rm{m^2}$. }
  \label{CBT100_SJT}
\end{figure}

The CBT consists altogether of 1000 junctions (ten parallel rows of 100 junctions in series) with an area of ($0.8\times0.8$) $\rm{\mu m^2}$ fabricated using optical lithography \cite{VTT}. Main advantage of optical lithography is that it reaches nowadays a sufficient resolution for sub micron structures and it allows the fabrication of numerous junctions with good yield and sufficient homogeneity. Figure \ref{CBT100_SJT} (top) shows three of these junctions with the attached cooling fins. Junction resistance is close to 10 k$\mathrm{\Omega}$ resulting in a total resistance on the order of 100 k$\mathrm{\Omega}$ for one sensor. One characteristic property for CBT is the capacitive charging energy ($E_{\rm{C}}/k_{\rm{B}}\approx 25$ $\rm{mK}$) determined by the junction size and the area of the island between junctions. 
Details of the SJT are depicted in Fig. \ref{CBT100_SJT} (bottom). The central junction used as a thermometer is embedded in four arrays of tunnel junctions ensuring on one hand the desired electromagnetic environment \cite{PRL-Pekola08} and allowing on the other hand a four probe measurement of the junction.

\section{Experimental setup}

We report a comparison of the thermometers with the realization of the PLTS-2000 \cite{JLTP-PLTS2000} at the PTB . A dilution refrigerator in a shielded room at PTB equipped with the $^3$He melting pressure thermometer (MPT) provide the platform for the experiment. A description of the high-accuracy PLTS-2000 realization at PTB including a detailed uncertainty budget is published elsewhere \cite{Engert07}. Both the investigated sensors are mounted in a copper case equipped with thermocoax lines for noise filtering \cite{Zorin} at the mixing chamber temperature. A superconducting coil surrounding the sensor setup provides the magnetic field to hold the sensors in the normal conducting state. 
\begin{figure}[htbp]
  \centering
    \includegraphics[width=0.99\textwidth]{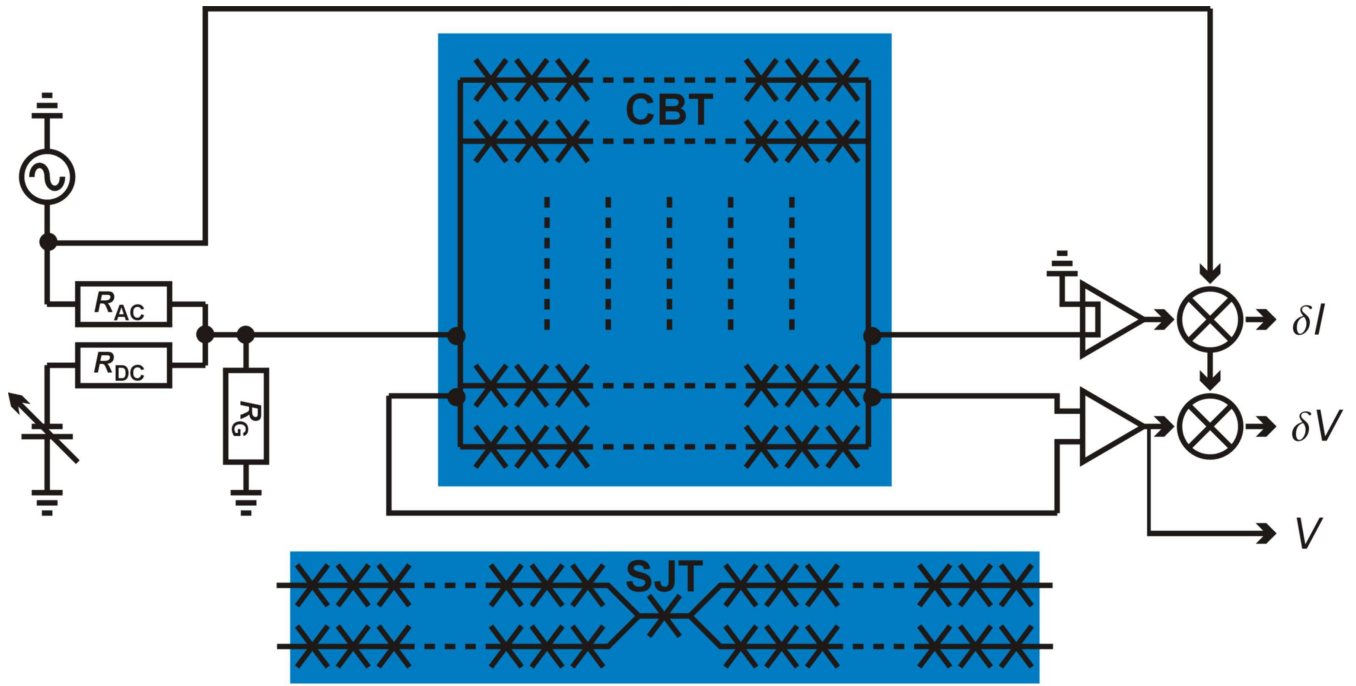}  
  
  \caption{Experimental setup for CBT measurements. The sensors at low temperature (blue areas) are connected with four wires to the room temperature electronics. Thermocoax cables filter the measurement lines at the mixing chamber temperature. EG\&G 5210 dual phase Lock-In-Amplifiers measure after a current preamplifier (DL instruments 1211) the AC component $\delta I$ and after a voltage preamplifier (DL instruments 1201) the $\delta V$ component. In addition, the amplified DC voltage component is measured with a digital multimeter. The SJT uses the identical setup: the applied voltage is approximately the same as in the CBT case due to the division in the array. The voltage across the central junction is measured with the second pair of arrays and an increased voltage amplifier gain.}
  \label{setup}
\end{figure}

Voltage is applied to the sample using room temperature resistive dividers, adding the separately applied DC and AC components. If $R_{\rm{G}} \ll R_{\rm{AC}}$ and $R_{\rm {G}} \ll R_{\rm{DC}}$, $V_{\rm{AC}}=V_{\rm{in}}\times R_{\rm{G}} / R_{\rm{AC}}$ and $V_{\rm{DC}}=V_{\rm{in}}\times R_{\rm{G}} / R_{\rm{DC}}$. Nominal values are (see Fig. \ref {setup}): $R_{\rm{G}}=47$ $\mathrm{\Omega}$, $R_{\rm{AC}}=100$ k$\mathrm{\Omega}$ and $R_{\rm{DC}}=10$ k$\mathrm{\Omega}$. In addition to applying a slowly swept DC voltage bias, an AC modulation of typically few tenths to few hundredths of Hz is added. The latter enables the measurement of the derivative of the $I$-$V$ curves using Lock-In amplifiers. The AC amplitude is set to be small (1\% to 3\%) as compared to the expected half width of the conduction curve.

\section{Comparison between the CBT sensor and the PLTS-2000}

A calibration of the DC voltage measurement is essential as the bias voltage ($V_{\rm{BIAS}}$) yields directly the measured temperature according to Eq. (\ref{CBTnumcorr}). Two methods are principally available for this task in case of CBT: either measuring the voltage including calibration of the voltage amplifier or calibration of the resistive voltage divider and voltage source yielding directly the applied voltage. The latter was used as the accuracy is higher and the resulting uncertainties are lower, even taking into account the neglected line resistance which was less than 0.1$\%$ of the sensor resistance. For the SJT measurements, only the direct measurement of the voltage with the voltage amplifier gain $10^4$ is suitable for the determination of the bias voltage of the single junction, as the sample divides the applied voltage by a factor of approximately 41, given by the number of junctions in the array plus the central junction. The junctions within the array are not similar enough for a precise determination of the division factor.

As $T_{\rm{CBT}}$ is derived from the normalized conductance curves the stability rather than the uncertainty of the conductance values is important. The resulting relative uncertainty for the normalized conductance values from the gain and output stabilities of the preamplifiers and the Lock-In amplifiers is estimated to be about $4\cdot10^{-4}$.

\begin{figure}[htbp]
  \centering
  \begin{minipage}[b]{0.46\textwidth}
    \includegraphics[width=0.97\textwidth]{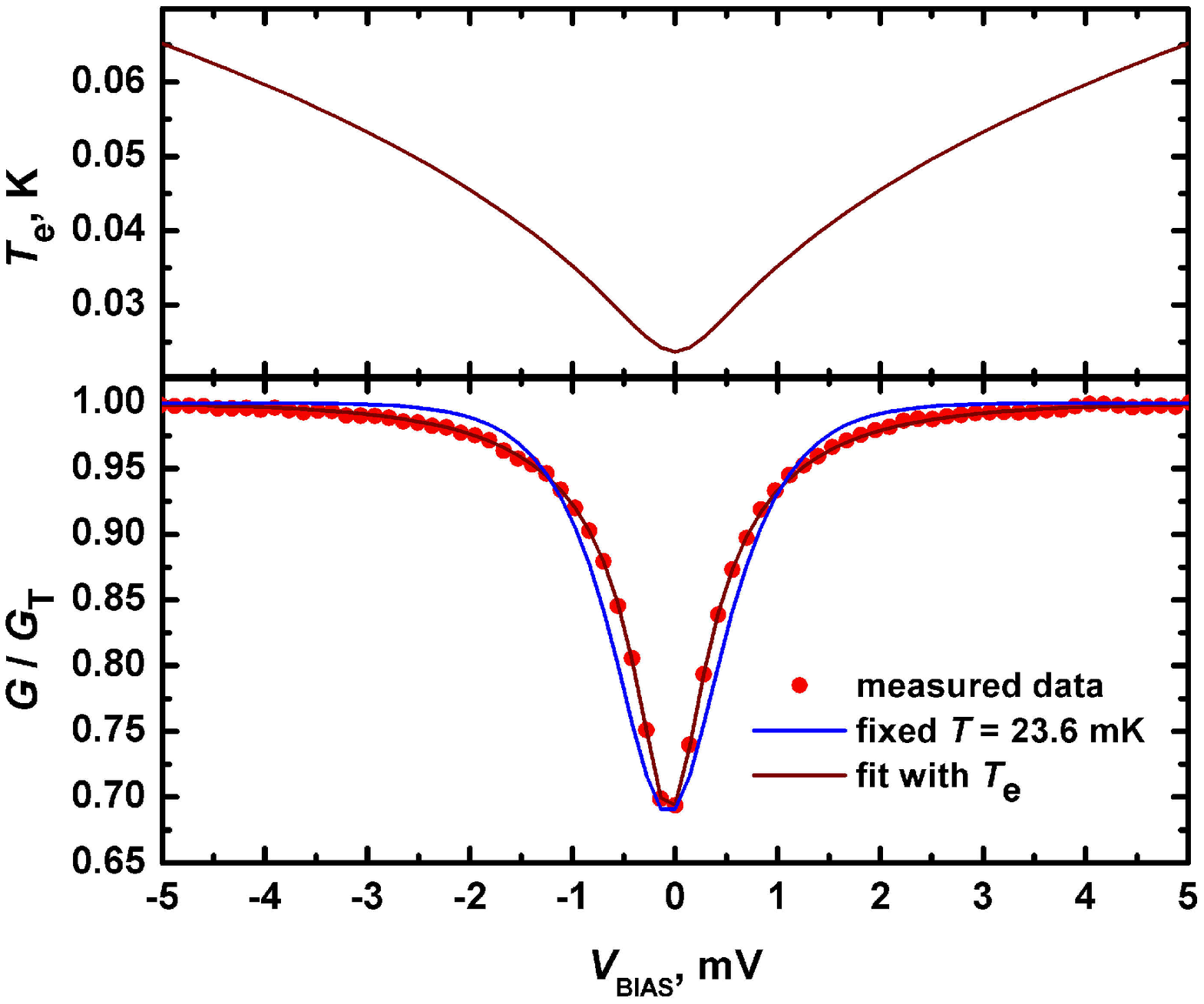}  
  \end{minipage}  
  \begin{minipage}[b]{0.49\textwidth}
    \includegraphics[width=0.95\textwidth]{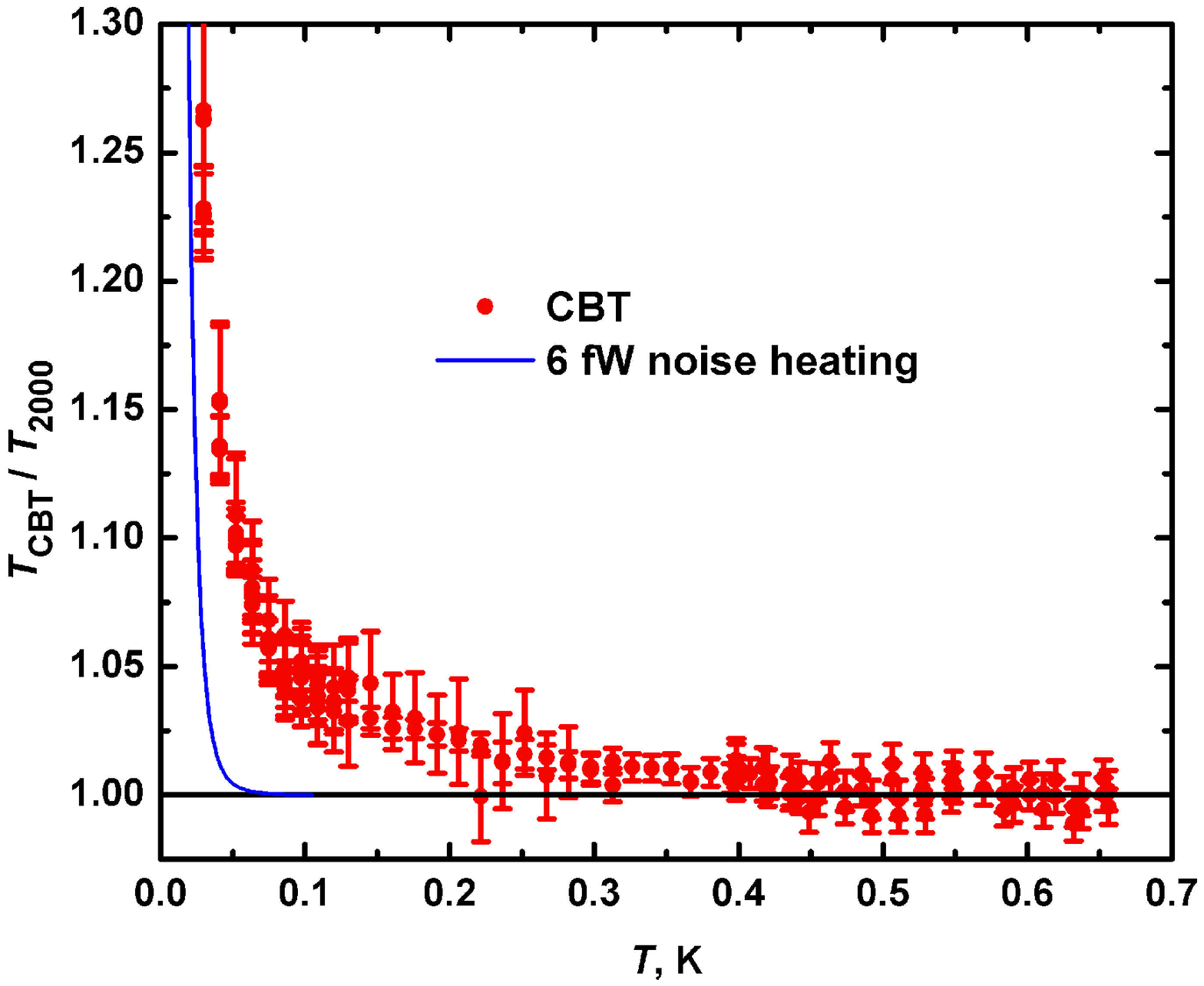}  
  \end{minipage}

  \caption{(left) Measured data of the CBT sensor at base temperature (red circles) of the cryostat and fit with the numerical model taking overheating effects into account. (top) The resulting electronic temperature as a function of applied bias voltage. (bottom) Measured data compared to two calculated curves: (blue) with fixed temperature of 23.6 mK, the electronic temperature at zero bias voltage, (dark red) electronic temperature follows the values of the top figure. (right) Deviation of the CBT temperature reading calculated using the numerical model from the PLTS-2000 temperature. The error bars represent the corresponding confidence interval for a coverage factor \textit{k}=1. The blue line depicts the expected deviation arising from a constant noise heating of 6 fW (see text). }
  \label{T26}
\end{figure}

A fit with the numerical model \cite{JLTP-Meschke} to one measurement (Fig. \ref{T26} (left)) at base temperature of the cryostat (T $\approx$ 0.008 K) determines all relevant sample parameters: the charging energy of $E_{\rm{C}}/k_{\rm{B}}$ $\equiv$ $e^2/2k_{\rm{B}}C$ = (27.5 $\pm$ 0.2) mK, the resistance of the sensor $R$ = (81.47 $\pm$ 0.02) k$\mathrm{\Omega}$ and the volume of the cooling fins of (3400 $\pm$ 250) $\mu \rm{m^3}$, assuming material parameter for the electron-phonon interaction of aluminum of $3\times10^8$ $\rm{WK^{-5}m^{-3}}$ , in good agreement with the fabricated volume of approx. 3000 $\mu \rm{m^3}$. The lowest observed electronic temperature is 23.5 mK, corresponding to a noise heating of approx. 6 fW.

Figure \ref{T26} (right) depicts an overview of the deviation of temperature from the CBT sensor with respect to the PLTS-2000 in the investigated temperature range from 0.008 K to 0.65 K. The agreement is good for the higher temperatures above 0.2 K, and stays within 5$\%$ down to $\leq$ 0.05 K. The solid line is calculated asuming a constant noise heat input of 6 fW and shows that the electron thermalization within the metal islands due to electron-phonon coupling is sufficient in the sensor design down to 0.05 K. Obviously, other thermal resistances lead to overheating of the sensor chip \cite{APL-Savin}, like the Kapitza resistance between the metal film and the silicon substrate. Imperfect glue between the silicon chip and the copper sample holder may begin to influence the result at the low temperature end.   

In the model \cite{cbt94,JLTP-Meschke} for the full description of the CBT, the temperature reading of the CBT enters the formulas only as a parameter and is derived from the experimental data as a result of a nonlinear approximation. For a correct uncertainty estimation for $T_{\rm{CBT}}$ we have applied Monte Carlo (MC) simulations \cite{GUM}. As an example, we focus on temperatures where the overheating effects are negligible and hence temperature can be derived using a direct fit without correction of the overheating effects. Figure 4 (left) demonstrates that overheating effects due to the electron-phonon coupling are negligible at temperatures above 0.4 K.
The procedure was as follows. For each conductance curve measured at defined reference temperatures we generated MC-samples by varying stochastically for every single data point the values for the normalized conductance as well as for the bias voltage within their individual uncertainties limits assuming a normal distribution of them. Then, to all those MC sample curves a non-linear fitting routine was applied to determine the individual MC-values $T_{\rm{CBT,MC}}$. As the resulting CBT temperature, $T_{\rm{CBT}}$, the mean of all $T_{\rm{CBT,MC}}$ values was taken, whereas the uncertainty $U$($T_{\rm{CBT}}$) was determined as the square root of the corresponding variance. Figure 4 (right) shows a comparison of the results for the MC calculation with uncertainty indications of a direct nonlinear fit for a the coverage factor \textit{k}=1. The uncertainty estimates for $T_{\rm{CBT}}$ from the MC calculations are more realistic as the uncertainties for the direct fit are obviously too small to account for the observed data point scattering. Contrary, at lower temperatures where the overheating effects were taken into account, the uncertainty estimates (Fig. 3 (right)) are mainly dominated by the large uncertainties of the above mentioned parameters entering the numerical model.

\begin{figure}[htbp]
  \centering
  \begin{minipage}[b]{0.49\textwidth}
    \includegraphics[width=\textwidth]{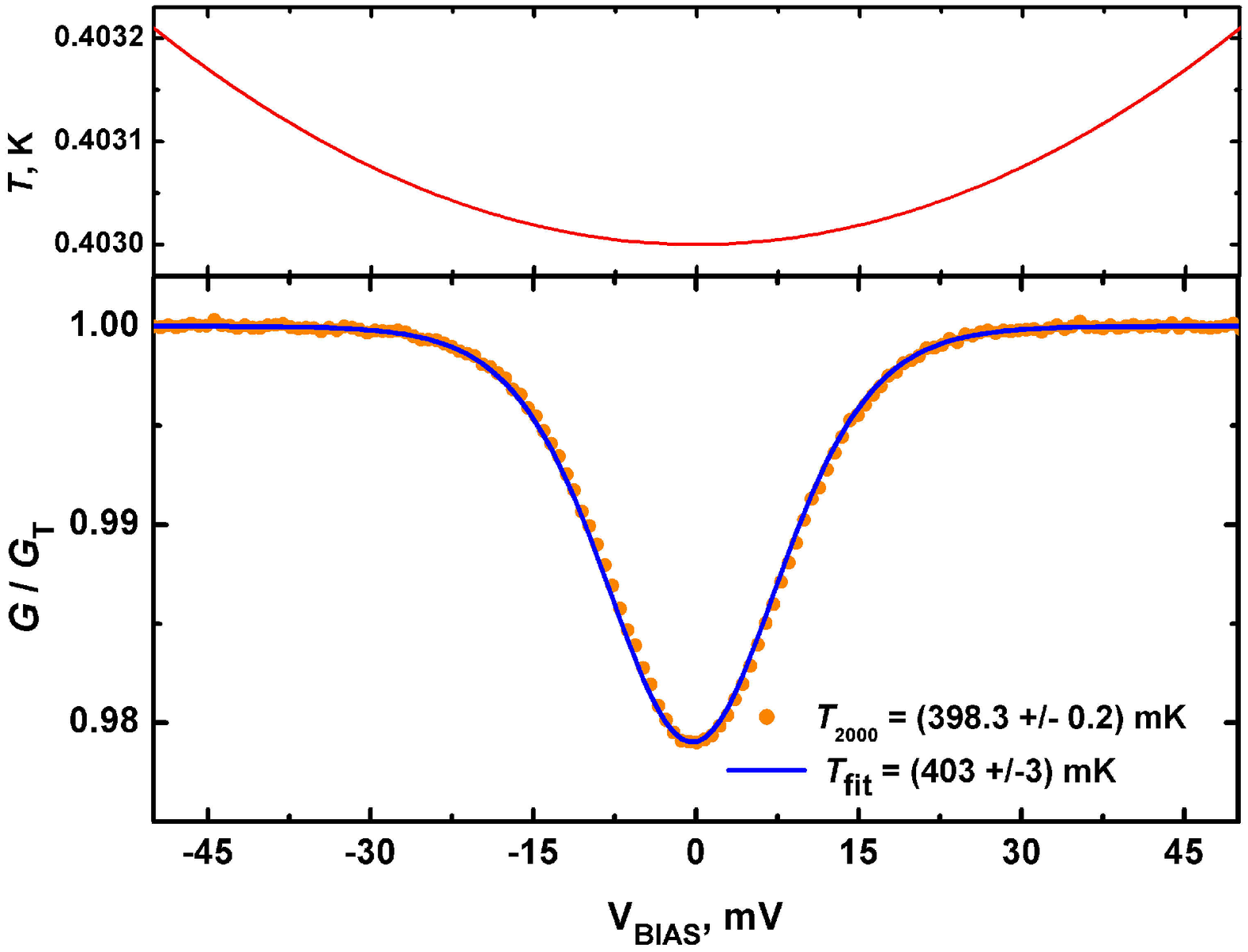}  
  \end{minipage}  
  \begin{minipage}[b]{0.49\textwidth}
    \includegraphics[width=\textwidth]{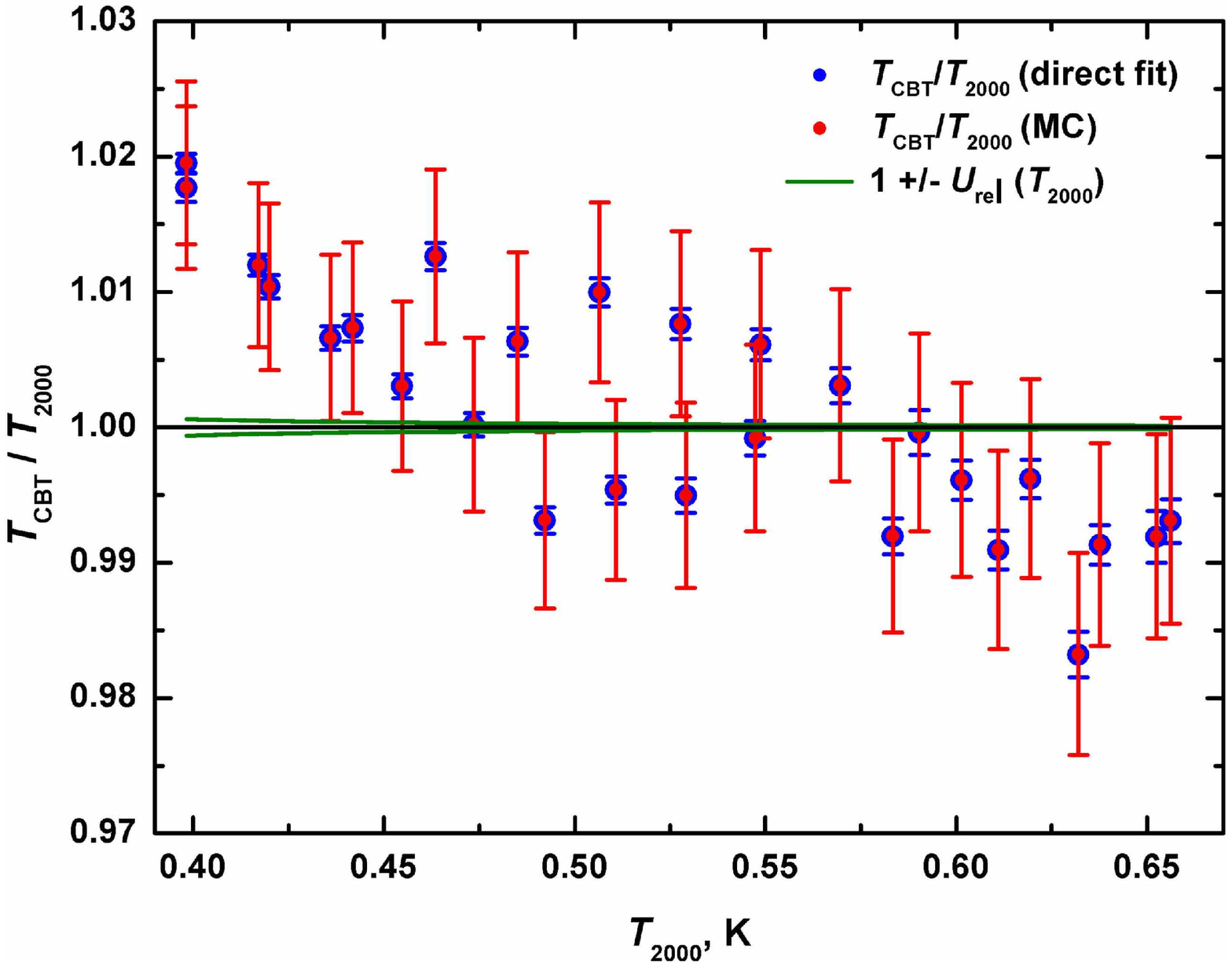}  
  \end{minipage}
  \caption{(left) One example of a CBT measurement (orange circles) at $T_{2000}$ = (398.3 $\pm$ 0.2) mK, together with the full model (blue line). The indicated confidence interval (\textit{k}=1) of $T_{\rm{fit}}$ (3 mK) are based on the fitting model to the measured data. The top panel depicts the increase of the electronic temperature due to applied bias voltage given by the thermal model, it uses the fixed island volume determined at lower temperatures (see Fig. \ref{T26}). (right) Direct comparison of the reference temperature ($T_{2000}$) with the temperature reading of the CBT sensor obtained from a direct fit. See text for explanation of the given uncertainty values.}
  \label{CBT_400}
\end{figure}
 \section{Comparison between the SJT sensor and the PLTS-2000}

\begin{figure}[htbp]
  \centering
  \begin{minipage}[b]{0.49\textwidth}
    \includegraphics[width=\textwidth]{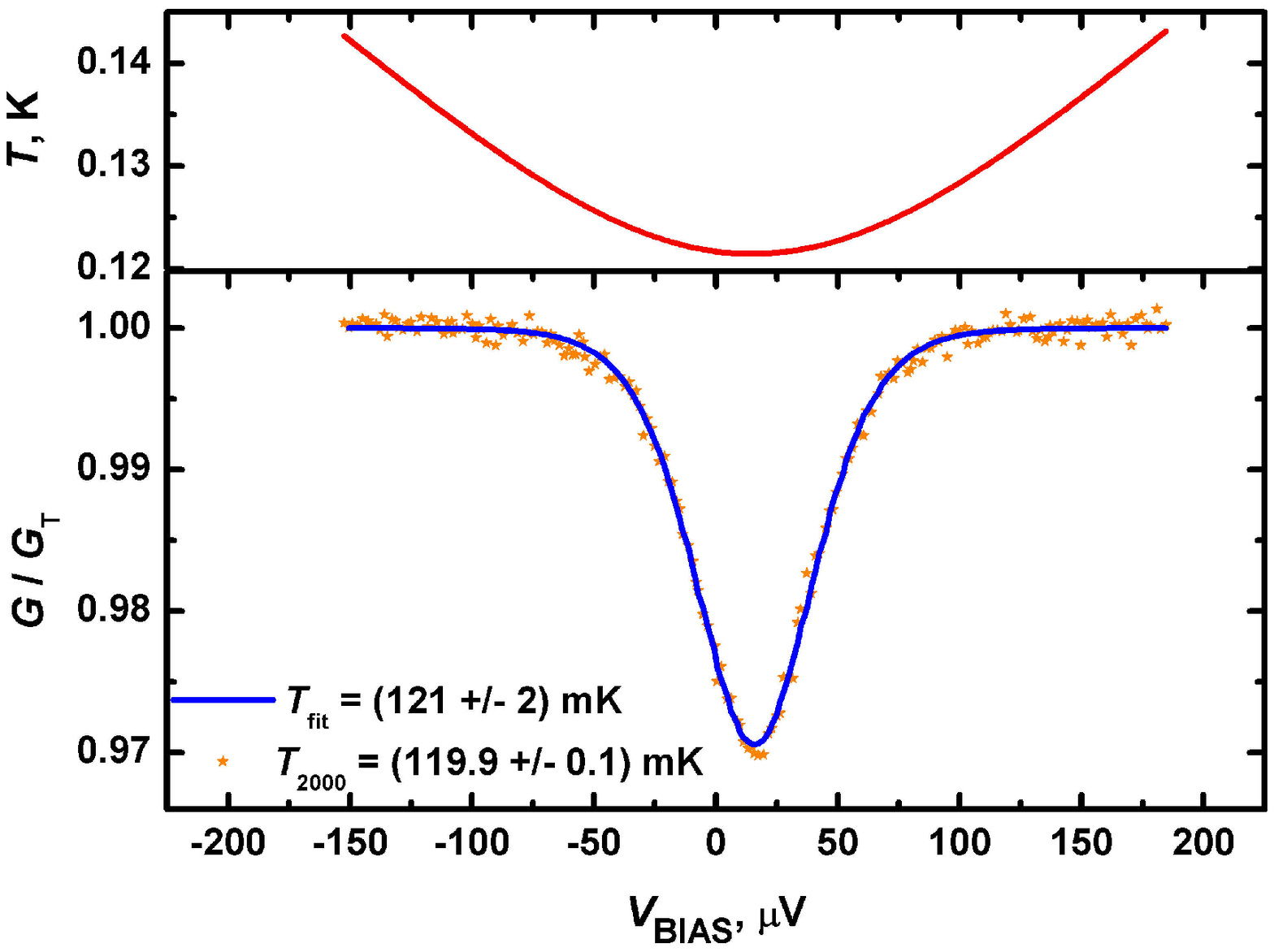}  
  \end{minipage}
  \begin{minipage}[b]{0.49\textwidth}
    \includegraphics[width=\textwidth]{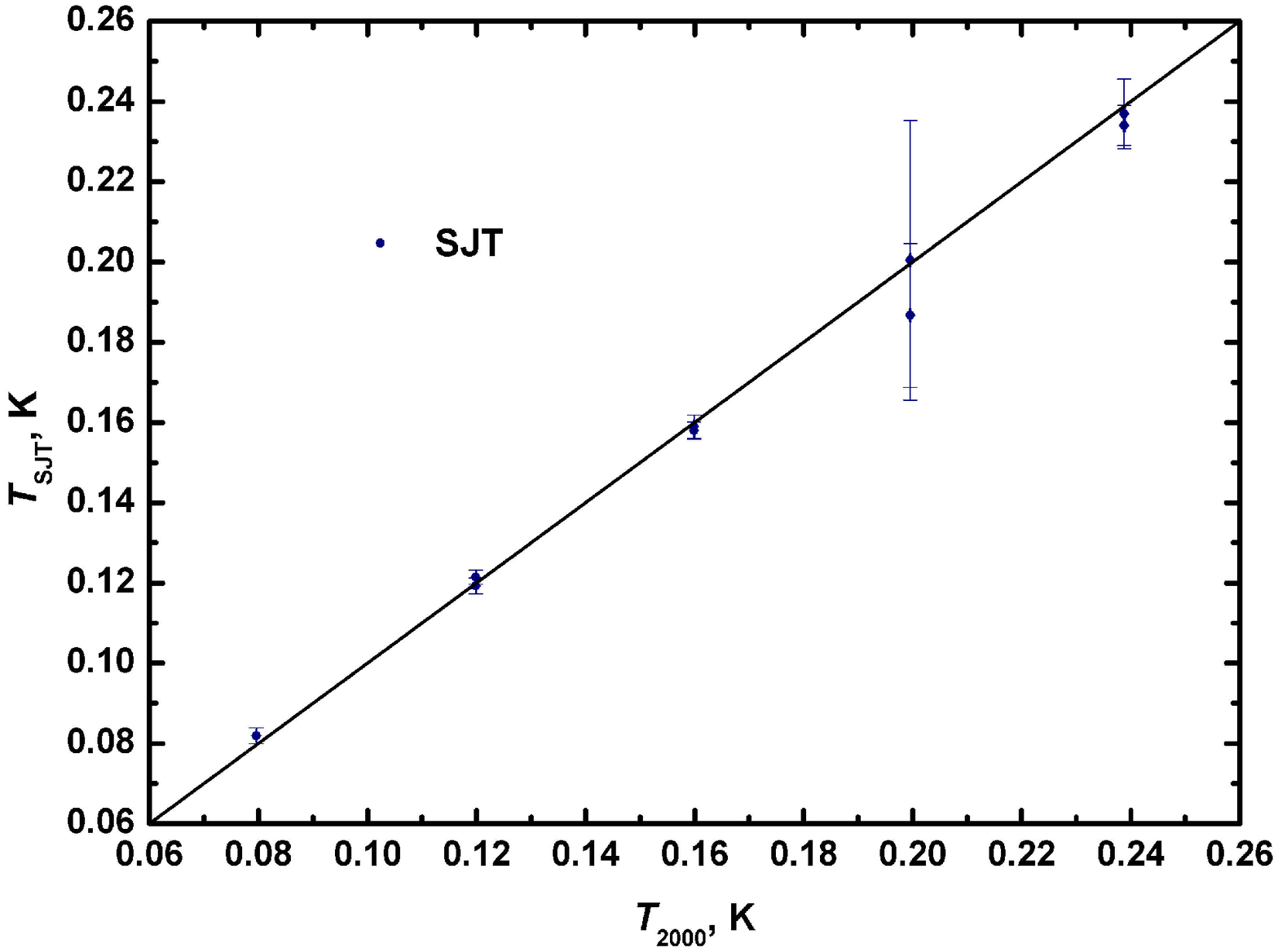}  
  \end{minipage}
  \caption{(left) A normalized SJT conduction curve as a function of applied bias voltage measured at $T_{2000}$ =  (119.9$\pm$0.1) mK. The fit with the model (blue line) is done with a constant island volume determined at a lower temperature resulting in the electronic temperature (red line) shown in the top panel. (right) Comparison between the temperature reading from SJT with the reference temperature. All SJT error bars depict the confidence interval (\textit{k}=1) of the fit to the data points. The difference in the accuracy at different temperatures is caused by the varying external measurement conditions.}
  \label{FINALSJT}
\end{figure}

In general, the measurement of the SJT requires an increased experimental effort compared to the CBT, as the peak width is quite small (on the order of 50 $\mu$V at temperatures around 0.1 K) in relation to the 100 times wider peak of the CBT. This can be compensated partly with longer time constants for the lock-in amplifiers and an increased measurement frequency for the AC modulation (from 27.7 Hz to 187 Hz). Still, the accessible temperature range is quite limited when the peak shrinks towards higher temperatures. Generally, the SJT temperature reading is in good agreement with the PLTS-2000 reference temperature. Figure \ref{FINALSJT} shows one experimental curve analyzed in the same way as the CBT and the result of this analysis at few temperatures. Even though the observed agreement of the SJT temperature reading with the reference temperature is satisfying, the uncertainty of the measurement leaves room for improvement.  

\section{Conclusion}

We demonstrated in this work that arrays of tunnel junctions produced with optical lithography are suitable for thermometry. We are able to present a detailed analysis of electron thermalization in the sensors due to the outstanding accuracy of the PLTS-2000 realization, resulting in valuable findings for future improvements of the sensor design: the fabrication homogeneity of the CBT was found to be sufficient for an accuracy of temperature reading on the order of few $\%$ whereas the thermalisation towards low temperatures still leave room for improvements. 
Moreover, we show in this paper the first experimental comparison of the SJT with the PLTS-2000. The temperature readings from the SJT agree within the achieved measurement accuracy at the reference temperature when overheating effects are included in the analysis. These results confirm the earlier experimental findings \cite{PRL-Pekola08} that an array of tunnel junctions provide a suitable environment for SJT. 
However, the resulting uncertainty for both thermometers is in the presented experiments still dominated by the measurement electronics at room temperature. Only a significant improvement in this respect would allow a systematic study of the remaining inherent errors of the devices itself. In future, a direct realization of the bias voltage using a Josephson voltage standard \cite{JVS} at low temperatures might allow a drastically increased readout accuracy for both sensors.
\begin{acknowledgments}
This work was supported in part by the European Community's Seventh Framework Programme (FP7/2007-2013) under grant agreement 228464 (MICROKELVIN)
\end{acknowledgments}

\end{document}